# Graph-Based Spatio-temporal Attention and Multi-Scale Fusion for Clinically Interpretable, High-Fidelity Fetal ECG Extraction

**Chang Wang**
Department of Electrical and
Computer Engineering
University of Nevada, Las Vegas
Las Vegas, USA
changw11@unlv.nevada.edu

**Ming Zhu**
Department of Electrical and
Computer Engineering
University of Nevada, Las Vegas
Las Vegas, USA
ming.zhu@unlv.edu

**Shengjie Zhai**
Department of Electrical and
Computer Engineering
University of Nevada, Las Vegas
Las Vegas, USA
shengjie.zhai@unlv.edu

**Buddhadeb Dawn**
Department of Internal Medicine
University of Nevada, Las Vegas
Las Vegas, USA
buddha.dawn@unlv.edu

**Shahram Latifi**
Department of Electrical and
Computer Engineering
University of Nevada, Las Vegas
Las Vegas, USA
shahram.latifi@unlv.edu

## ABSTRACT

Congenital Heart Disease (CHD) is the most common neonatal anomaly, highlighting the urgent need for early detection to improve outcomes. Yet, fetal ECG (fECG) signals in abdominal ECG (aECG) are often masked by maternal ECG and noise, challenging conventional methods under low signal-to-noise ratio (SNR) conditions. We propose **FetalHealthNet (FHNet)**, a deep learning framework that integrates Graph Neural Networks with a multi-scale enhanced transformer to dynamically model spatio-temporal inter-lead correlations and extract clean fECG signals. On benchmark aECG datasets, FHNet consistently outperforms long short-term memory (LSTM) models, standard transformers, and state-of-the-art models, achieving $R^2>0.99$ and RMSE $\approx 0.015$ even under severe noise. Interpretability analyses highlight physiologically meaningful temporal and lead contributions, supporting model transparency and clinical trust. FHNet illustrates the potential of AI-driven modeling to advance fetal monitoring and enable early CHD screening, underscoring the transformative impact of next-generation biomedical signal processing.

## 1 Introduction

Congenital Heart Disease (CHD), the most common structural anomaly in newborns, affects 0.8%–1.2% of live births worldwide and remains a leading cause of neonatal mortality and long-term morbidity. [9] Early detection is essential; however, prenatal echocardiography — the gold standard — is limited by fetal positioning, amniotic fluid variability, maternal body composition, and operator dependency. These constraints can obscure subtle anomalies and delay timely diagnosis. Maternal abdominal electrocardiography (aECG) provides a compelling, non-invasive, and portable alternative for continuous fetal cardiac monitoring. Unlike echocardiography, which offers only a snapshot, aECG records dynamic cardiac activity, enabling detection of transient events such as arrhythmias and supporting broader screening in resource-limited settings. The key challenge lies in extracting the weak fetal electrocardiogram (fECG) signal, which is overshadowed by maternal ECG (mECG) and contaminated by physiological and environmental noise. [3] Recent advances in deep learning, particularly transformer architectures and Graph Neural Networks (GNNs), offer new solutions for low-SNR signal extraction. Transformers leverage self- and cross-attention to isolate weak signals from noisy sequences, while GNNs capture inter-lead and spatio-temporal dependencies (e.g., DSTAGNN [12]). Building on these insights, we propose **FetalHealthNet (FHNet)**, a graph-structured, multi-scale enhanced transformer model integrating GNN-based inter-lead modeling with spatio-temporal and multi-scale attention to achieve high-fidelity fECG extraction and reliable CHD risk screening This study significantly promotes the robustness and quality of fECG extraction. The primary contributions are four-folded:

- FHNet. A graph-structured, multi-scale transformer tailored for precise and robust fECG extraction and CHD risk screening from multi-lead aECG under low SNR.
- Flexible graph construction and GNN modules to capture time-varying inter-lead correlations to improve fidelity of the extracted fECG under dynamic environment.
- Both encoders and decoders are coupled with multi-scale spatio-temporal cross-attention layers to enhance the accuracy of fECG extraction ($R^2 > 0.99$ and RMSE ≈ 0.015).
- Clinical and data interpretability. Attention visualizations and Integrated Gradients expose key temporal/spatial dependencies, supporting clinical transparency and acceptance.

By systematically addressing these challenges, the proposed FHNet not only bridges cutting-edge deep learning techniques with pressing clinical needs but also offers a scalable and interpretable solution for early CHD detection, with the potential to significantly improve neonatal outcomes.

## 2  Literature Work

Accurate fetal cardiac assessment has driven decades of research in electronic fetal monitoring (EFM). Cardiotocography (CTG) is widely used for FHR but offers time-averaged trends and lacks beat-to-beat morphology for early evaluation of conditions such as CHD [4]. Fetal magnetocardiography (FMCG) provides high fidelity yet is costly and impractical [17], while scalp ECG (SECG) is invasive and limited to intrapartum care. Non-invasive fetal ECG (fECG) from maternal abdominal ECG (aECG) therefore emerges as a route to continuous, accessible, morphology-level monitoring.

Clinical deployment of fECG is constrained by severe signal degradation. The fetal signal is typically 10–50× weaker than maternal ECG (mECG) and is further contaminated by maternal electromyographic activity, fetal motion, uterine contractions, and environmental artifacts [6]. These interferences overlap with fECG in both time and frequency domains, making separation fundamentally challenging. Early approaches bifurcated into temporal and spatial strategies. Temporal methods (e.g., template subtraction and Kalman filtering) operate on single-channel recordings to estimate and remove mECG, but performance hinges on template accuracy, which can vary within minutes as physiology changes [10,16]. Spatial methods, notably blind source separation (BSS) such as Independent Component Analysis (ICA), exploit multi-channel aECG to statistically disentangle sources; however, they often require many electrodes—limiting portability—and degrade markedly at low signal-to-noise ratios (SNR) and under non-stationarity [1,2,13,14,19]. Neither class consistently preserves fetal waveform morphology when maternal/fetal complexes temporally overlap.

Deep learning has reshaped the field by learning task-specific representations directly from raw signals. Convolutional neural networks (CNNs) capture local morphology (e.g., QRS, ST-T) but struggle with long-range dependencies; recurrent models such as recurrent neural networks (RNNs) and long short-term memory (LSTM) models encode temporal context but can be computationally heavy and prone to vanishing gradients in long sequences [11,23]. The transformer architecture addresses these limitations via self-attention, enabling efficient modeling of long-range dependencies independent of sequence distance while retaining local detail through multi-head mechanisms [18]. Building on this, attention-based generative and discriminative models have improved fECG extraction in noisy conditions. For example, integrating attention with CycleGAN has enhanced waveform reconstruction under domain shifts [15]; PA²Net explicitly models ECG periodicity to stabilize detection [21]; and CAA-CycleGAN introduces correlation-aware attention to mitigate noise correlated across channels [22]. While promising, most of these approaches primarily emphasize temporal dynamics and under-utilize the rich spatial structure inherent to multi-lead aECG.

Multi-lead aECG provides distinct projections of the fetal cardiac field; jointly modeling cross-lead relationships is therefore crucial at low SNR. GNNs offer a principled way to encode inter-lead dependencies by representing each lead as a node with edges reflecting physiological geometry or learned statistical couplings [5,7]. CSGSA-Net, for instance, introduces a graph-based sparse attention mechanism to better distinguish fetal from maternal sources, but limited multi-scale fusion and generative capacity constrain waveform fidelity [20]. Beyond fECG, dynamic spatio-temporal graph architectures (e.g., DSTAGNN) have proven effective at capturing evolving spatial correlations over time in other domains, underscoring the value of adaptive graphs that adjust to context [12]. However, these ideas have not been fully adapted to high-fidelity, generative fECG reconstruction where both cross-lead (spatial) and cross-scale (temporal) cues must be integrated end-to-end.

Collectively, the literature reveals a clear gap: the absence of a unified, generative framework that (i) dynamically models inter-lead spatial structure, (ii) captures multi-scale temporal dependencies, and (iii) preserves clinically salient morphology under low-SNR and non-stationary conditions. The complementary strengths of transformers (long-range temporal reasoning) and GNNs (explicit spatial relational modeling) provide a compelling path forward. To this end, we propose **FHNet**, a unified encoder–decoder that couples dynamic spatiotemporal graph modeling with a multi-scale, attention-enhanced transformer. The encoder builds adaptive inter-lead graphs with spatial/temporal attention; the decoder performs multi-scale fusion and cross-attention to reconstruct morphology-faithful fECG. Built-in interpretability (attention visualization and attribution) supports clinician trust. FHNet delivers robust, clinically transparent extraction suitable for early CHD screening and scalable fetal rhythm surveillance.

## 3  Methodology

This study presents FetalHealthNet (FHNet), a novel encoder-decoder deep-learning architecture for high-precision extraction of fECG signals from multi-channel aECG recordings and potentially downstream clinical applications such as early detection of fetal CHD. The encoder incorporates with DSTAGNN [12] principles to model complex spatio-temporal dependencies in multi-lead aECG signals, while the decoder employs an enhanced self-attention mechanism to reconstruct high-fidelity fECG signals through dynamic contextual integration. Multi-Scale Temporal Feature Extraction (MSTFE) modules are embedded in both encoders and decoders to capture signal characteristics across multiple temporal scales, enhancing robustness to diverse clinical conditions.

## 3.1 Symbol Definitions and Problem Formulation

Denote the multi-channel aECG input as $X_{aECG} \in \mathbb{R}^{B \times L_m \times 1 \times T_{in}}$, where $B$ is batch size, $L_m$ represents the number of abdominal leads (graph nodes), and $T_{in}$ is the input time length. The initial feature dimension for each lead at each time point is assumed to be 1. The target fECG output is $Y_{fECG} \in \mathbb{R}^{B \times T_{out} \times D_y}$, with $D_y = 1$ representing the single-lead fECG signal. Given the multi-lead structural information $G$ (implicit and/or explicit), the model $F$ estimates $Y_{fECG} = F(X_{aECG}; G)$.

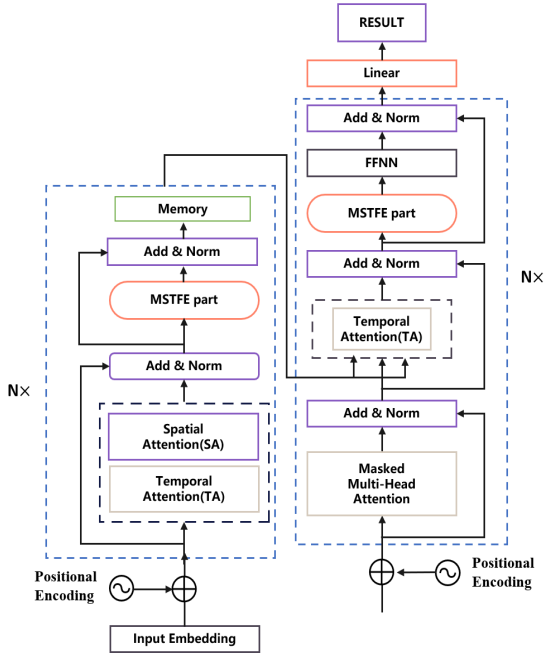

Figure 1: FHNet encoder–decoder framework architecture

## 3.2 FHNet Network Architecture

The FHNet architecture (Figure 1) employs a scalable encoder-decoder framework with enhancement for dynamic multi-scale spatio-temporal feature extraction from multi-channel aECG recordings. The modular design enables flexible depth configuration to accommodate varying signal complexity and clinical requirements. The encoder part comprises N stacked encoder blocks, capturing spatio-temporal dependencies via sequential temporal attention (TA), spatial attention (SA)-modulated Graph Convolution, and MSTFE modules. The decoder part is also N-stacked, and each integrates TA and MSTFE modules for improved temporal resolution, but omits the SA for single-lead fECG outputs.

*3.2.1 Encoder Module with Multi-Scale Memory Propagation.* Each encoder block is composed of a TA, a SA, and a MSTFE module (Figure 1) to learn rich spatio-temporal representations from multi-lead aECG signals. The TA is a multi-head temporal self-attention for the multi-lead aECG input sample $\mathcal{X}'(l)$, which is the input to the $l$-th encoder layer. The query $(Q^{(l)})$, key $(K^{(l)})$, and value $(V^{(l)})$ of TA are computed as:

$$Q^{(l)} \triangleq \mathcal{X}'^{(l)} W_q^{(l)}, \quad K^{(l)} \triangleq \mathcal{X}'^{(l)} W_k^{(l)}, \quad V^{(l)} \triangleq \mathcal{X}'^{(l)} W_v^{(l)} \quad (1)$$

where $W_q^{(l)}, W_k^{(l)}, W_v^{(l)}$ are trainable weights. The TA output is:

$$Att(Q^{(l)}, K^{(l)}, V^{(l)}) = Softmax(A^{(l)}) V^{(l)} \quad (2)$$

where $A^{(l)} = \frac{Q^{(l)} K^{(l)\top}}{\sqrt{d_h}} + A^{(l-1)}$.

For H attention heads:

$$O^{(h)} = Att\left(Q W_q^{(h)}, K W_k^{(h)}, V W_v^{(h)}\right), h = 1, 2, \ldots, H \quad (3)$$

$$O = [O^{(1)}, O^{(2)}, \ldots, O^{(H)}] \quad (4)$$

Then, the residual attention mechanism is employed by adding the current module attention matrix $Att^{(l)}$ with the one from the previous module $Att^{(l-1)}$. This helps the model fuse temporal dependency information from different depths and alleviates the vanishing gradient problem. After being concatenated, projected, reshaped back to the original tensor structure and a normalization layer (LayerNorm), the TA output yields

$$Y_{TA} = \text{LayerNorm}(\text{Linear}(\text{Reshape}(O) + \chi')) \quad (5)$$

where $Y_{TA} \in \mathbb{R}^{B \times L_m \times D' \times T_{curr}}$.

$Y_{TA}$ is then fed into the Spatial Attention module to learn dynamic spatial dependencies between different aECG leads. An improved multi-head self-attention mechanism computes spatial attention scores $S_{SAt} \in \mathbb{R}^{B \times K_{cheb} \times L_m \times L_m}$ (where $K_{cheb}$ is the order of the Chebyshev graph convolution, also serving as the number of heads for SA). Unlike standard self-attention with direct weight vectors, the output of SA ($S_{SAt}$) primarily serves as a dynamic weight to modulate inter-lead information aggregation in the subsequent graph convolution module.

Meanwhile, graph convolution is applied based on Chebyshev polynomial approximation to aggregate information from neighbor leads, thereby extracting structure-aware spatial features. A baseline adjacency information is determined by the static predefined connected adjacency, while a normalized Laplacian matrix $L$ is generated based on the adjacency matrix (defaulting to an identity matrix) to calculate the Chebyshev polynomials. Then, a learnable dynamic graph structure adjustment is applied by introducing a learnable mask $M_{learn}^{(k)} \in \mathbb{R}^{L_m \times L_m}$ for each Chebyshev order k in the graph convolution. This mask is element-wise multiplied with the graph derived from the baseline adjacency information, forming a dynamically adjusted adjacency component:

$$A_{\text{dyn\_comp}}^{(k)} = adj_{\text{pa\_static}} \odot M_{\text{learn}}^{(k)}. \quad (6)$$

This dynamic component is then added to the spatial attention scores $S_{SAt}$, yielding an effective dynamic spatial dependency factor $P_{\text{eff}}^{(k)}$ after a softmax layer:

$$P_{\text{eff}}^{(k)} = \text{Softmax}\left(S_{\text{SAt}}^{(k)} + A_{\text{dyn\_comp}}^{(k)}\right). \quad (7)$$

Crucially, $P_{eff}^{(k)}$ can element-wise modulate each term of the Chebyshev polynomial $T_k(L)$. This allows the graph's connection strengths and structure to be adaptively learned end-to-end based on aECG data characteristics, enabling the model to more flexibly capture complex spatial dependencies between leads, surpassing the limitations of fixed graph structures. The graph convolution operation at each time point $t$ can be represented as:

$$H_{GCN}(t) = \text{ReLU}\left(\sum_{k=0}^{K_{cheb}-1} \theta_k \left(T_k(L) \odot P_{eff}^{(k)}\right) Y_{TA}\right), \quad (8)$$

where $\theta_k$ are learnable coefficients and $T_k(L)$ is the $k$-th order Chebyshev polynomial, and $H_{GCN} \in \mathbb{R}^{B \times L_m \times D_{chev\_filt} \times T_{curr}}$.

In the end, we integrate a MSTFE module via parallel Gated Tanh Units (GTUs) to further refine temporal dynamics from the spatially fused features $H_{GCN}$, particularly to capture features of mECG and weak fECG signals at different time scales. In this work, $H_{GCN}$ (with adjusted dimension order) is fed into three parallel GTU branches, each with different 1D convolution kernel sizes (e.g., 3, 5, 7, respectively) and the 1D convolution within each individual branch has a different receptive field (e.g., kernel sizes of 3,5, and 7 MSTFE in the Encoder (with GTUs): To refine temporal dynamics, $H_{GCN}$ is processed by an MSTFE module comprising three parallel Gated Tanh Units (GTUs) with different 1D convolution kernel sizes (e.g., 3, 5, 7):

$$GTU(Z) = \tanh(Z_E) \odot \sigma(Z_F), \quad (9)$$

where Z is the input to the GTU branch, $Z_E$ and $Z_F$ are the two halves of the input split along the channel dimension after their respective convolutional layers, where $\sigma$ is the Sigmoid function. Outputs from each branch are then concatenated and projected to yield $H_{\text{MSTFE\_enc}}$.

Finally, the output of the encoder's MSTFE, $H_{\text{MSTFE\_enc}}$, is added to a residual signal derived from an earlier stage of the module, and is passed through a ReLU activation and Layer Normalization (Layer Norm) to form the final output of the current module, $X_{\text{block\_out}} \in \mathbb{R}^{B \times L_m \times D_{chev\_filt} \times T_{next}}$. The accumulated attention scores $A^{(l)}$ are also passed to the next layer.

*3.2.2 Encoder Memory Generation.* The outputs $\{X_{\text{block\_out}}^{(i)}\}_{i=1}^{N}$ from all $N$ encoder blocks are concatenated along the time dimension, forming $H_{aggregated} \in \mathbb{R}^{B \times L_m \times D_{chev\_filt} \times T_{total\_concat}}$. This tensor $H_{aggregated}$ undergoes dimension permutation and reshaping: the lead dimension $L_m$ and feature dimension $D_{chev\_filt}$ are merged, ultimately achieving the memory tensor $M_{enc} \in \mathbb{R}^{B \times T_{total\_concat} \times (L_m \cdot D_{chev\_filt})}$, which serves as the encoder's final contextual representation of the input aECG signal and is passed to the decoder, fulfilling the "memory" function.

*3.2.3 Multi-Scale Enhanced Decoder for fECG Generation.* The decoder autoregressively generates the target fECG sequence $Y_{fECG}$ based on $M_{enc}$ and previously generated fECG segments. The decoder input is linearly embedded and augmented with sinusoidal positional encoding to obtain $H_{emb} \in \mathbb{R}^{B \times T_{out} \times D_{model}}$.

To better capture subtle patterns at different time scales when generating fECG (such as morphological changes in fQRS waves), an MSTFE module is also employed after the decoder input embedding (but without GTUs). It also consists of multiple parallel 1D convolutional layers, each with a different kernel size (e.g., 3, 5, 7). These convolutions operate directly on the time dimension and use ReLU as the activation function:

$$H_{\text{MSTFE\_dec}} = \|_{r \in 3,5,7} \text{ReLU}(\text{Conv1D}_r(H_{emb})). \quad (10)$$

The outputs of the 3 branches are concatenated along the feature dimension, and its total output dimension remains $D_{model}$. The output of this MSTFE module, $H_{\text{MSTFE\_dec}}$, is added to the original $H_{\text{dec\_emb}}$ via a residual connection and goes through normalization:

$$H'_{emb} = \text{LayerNorm}\left(H_{emb} + \text{Proj}(H_{\text{MSTFE\_dec}})\right). \quad (11)$$

$H'_{emb}$ is then fed to the subsequent transformer decoder layer, and so on, until the last decoder layer. which consist of Masked Multi-Head Self-Attention, Cross-Attention (querying $M_{enc}$), and Feed-Forward Networks (FFNN). A final linear layer maps the hidden states to $D_y$ to produce $Y_{fECG}$.

## 4 Data Processing and Experiment Results

### 4.1 Data Processing

The proposed FHNet was examined on two publicly available datasets, namely Abdominal and Direct Fetal ECG Dataset (ADFECGDB) [8] and Fetal ECG Synthetic Dataset (FECGSYNDB) [24], to ensure both clinical relevance and robust validation, together with other baseline models and state-of-the-art model CSGSA-Net [21] for performance comparisons. The ADFECGDB [8] served as a real-world benchmark, including independent 5-minute multi-channel recordings from five laboring pregnant participants at term (gestational age 38–41 weeks). Meanwhile, FECGSYNDB [24] contains 1,750 unique recordings from 10 simulated subjects across various scenarios, including uterine contractions and a range of signal-to-noise ratios. A unified data preprocessing pipeline was applied to all recordings for consistency. Baseline wander was removed using a dual-pass median filter, followed by high-frequency noise suppression via a sixth-level Discrete Wavelet Transform (DWT).

All models were implemented in PyTorch and trained on NVIDIA GeForce RTX 3080 Ti GPU (7,424 CUDA cores, 16 GB VRAM). To ensure fair comparison, key hyperparameters were standardized: training was performed with the Adam optimizer (initial learning rate = 1e$^{-4}$) and a batch size of 32. The specific number of training epochs and sliding-window configurations for each experiment are provided in the corresponding result tables.

### 4.2 Experimental Results

A side-by-side model performance comparison on ADFECGDB and FECGSYNDB is shown in Table 1. As one can see, FHNet variants consistently surpassed all baseline models in $R^2$ (> 0.99) and compatible RMSE, if not better, for both ADFECGDB and FECGSYNDB, using the comparable encoder-decoder window sizes and training epochs. The outcomes reflect the combined effect of three critical architectural choices: the integration of graph-based spatial attention in the encoder to model inter-lead dependencies, the use of multi-scale temporal feature extraction in the decoder to handle temporal patterns at different resolutions, and the careful alignment of encoder and decoder capacities to ensure that extracted features are fully utilized during reconstruction. The near-perfect $R^2$ score indicates that FHNet preserves the fECG waveform morphology with a fidelity suitable for clinical applications, including accurate fetal heart rate measurement and morphological anomaly detection. As illustrated in Figure 2, the reconstructed QRS complexes closely match the ground truth, maintaining precise amplitude and slope characteristics. This attribute is essential for reliable clinical interpretation, as it ensures diagnostic features remain intact rather than being distorted by reconstruction. The consistent results between the real-world and synthetic datasets, underscore that FHNet offers not only superior accuracy but also robust and reproducible performance across markedly different acquisition scenarios.

**Table 1: Performance Comparison experiments with Baseline Models on the ADFECGDB and FECGSYNDB dataset**

| Models | ADFECGDB | | FECGSYNDB | | Train Epoch | Window size |
|---|---|---|---|---|---|---|
| | $R^2$ | RMSE | $R^2$ | RMSE | | |
| LSTM | 0.921 | 0.0926 | 0.990 | 0.0161 | 30 | 300-50 |
| Transformer | 0.973 | 0.1240 | 0.948 | 0.0526 | 30 | 300-50 |
| FHNet_1 | 0.981 | 0.0244 | 0.978 | 0.0238 | 30 | 100-50 |
| FHNet_2 | 0.991 | 0.0221 | 0.991 | 0.0157 | 30 | 300-50 |
| FHNet_3 | 0.994 | 0.0191 | 0.996 | 0.0112 | 30 | 500-50 |
| CSGSA-Net_1 | 0.927 | 0.0151 | 0.951 | 0.0110 | 500 | 128-128 |
| CSGSA-Net_2 | 0.691 | 0.0434 | 0.942 | 0.0311 | 500 | 300-300 |
| FHNet_4 | 0.995 | 0.0155 | 0.975 | 0.0255 | 30 | 100-100 |
| FHNet_5 | 0.991 | 0.0252 | 0.964 | 0.0308 | 30 | 300-300 |

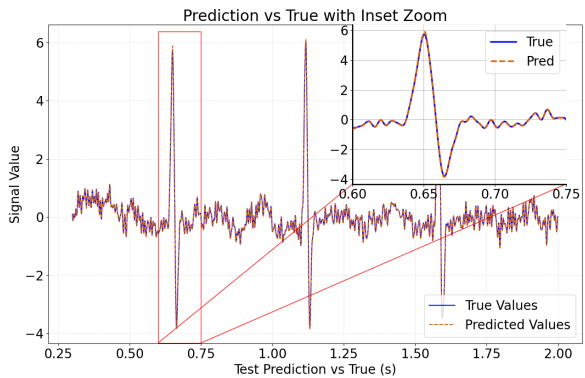

**Figure 2: fECG reconstruction with FHNet. Left: full-sequence waveform with QRS region highlighted. Right: zoomed view showing high-fidelity reproduction of peaks, troughs, and morphology.**

## 4.3 Interpretability Analysis

To further demonstrate the interpretability of our proposed FHNet model, a 2-second multi-lead aECG segment with corresponding fECG signal were analyzed using **Integrated Gradients (IG)** and **internal spatial attention (SAt)** to quantify the spatio-temporal contributions to fECG signal reconstruction/extraction.

**Table 2: Normalized IG attribution fraction for aECG leads within a 0.25 second window**

| | 0.00s-0.25s | 0.25s-0.50s | 0.50s-0.75s | 0.75s-1.00s |
|---|---|---|---|---|
| Lead_1 | 0.5558 | 0.8692 | 1.0000 | 1.0000 |
| Lead_2 | 1.0000 | 0.5219 | 0.6150 | 0.4093 |
| Lead_3 | 0.0000 | 0.0000 | 0.0000 | 0.0000 |
| Lead_4 | 0.5871 | 1.0000 | 0.1095 | 0.8157 |
| | 1.00s-1.25s | 1.25s-1.50s | 1.50s-1.75s | 1.75s-2.00s |
| Lead_1 | 0.8208 | 0.8970 | 0.0000 | 1.0000 |
| Lead_2 | 0.6877 | 0.6427 | 1.0000 | 0.5044 |
| Lead_3 | 0.0000 | 0.0000 | 0.0549 | 0.0000 |
| Lead_4 | 1.0000 | 1.0000 | 0.5380 | 0.9891 |

*4.3.1 Integrated Gradients (IG).* IG quantifies the importance of each input feature by calculating the integral of gradients of the model's specific output (e.g., the predicted fECG signal sequence) with respect to the input features (in this context, the signal values of different aECG leads at various time points). Table 2 presents the obvious heterogeneity of each lead contributions that Lead 1, 2, and 4 contribute significantly in the corresponding fECG extraction over the 2 seconds, whereas Lead 3's contribution is almost negligible. On the other hand, the importance of the main contributing leads (i.e., Lead_1, _2, _4) is not constant but shows dynamic changes over the 2-second input signal period, demonstrating the needs of dynamic multi-scale spatio-temporal attention on the delicate multi-lead aECG processing.

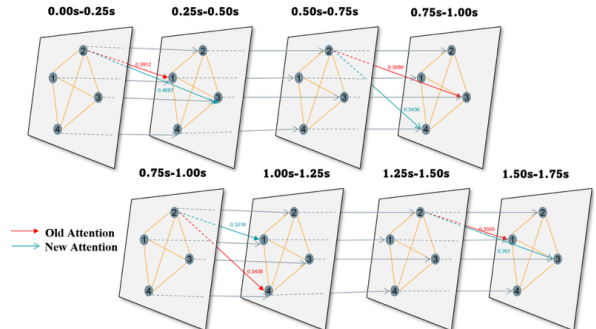

**Figure 3: Attention transfer diagram of different leads in prediction data**

*4.3.2 Attention Mechanisms.* Within FHNet, SAt provides data-driven inter-lead weighting that modulates subsequent graph aggregation. Examination of SAt trajectories (Figure 3) reveals the attention shifts among informative leads that mirror the IG profiles. Notably, a substantial transfer value (≈ 0.4246) accompanies a shift in Lead 2's primary affinity from Lead 1 to Lead 3, suggesting adaptive re-weighting in response to time-varying signal quality and/or salient features.

Through the joint analysis of IG and SAt, this study not only confirms the fECG extraction performance of FHNet, but also more crucially, significantly enhances the interpretability of the model's internal decision-making mechanisms. This approach provides an in-depth understanding on how the model cooperates with multi-lead signals, and how dynamic inter-lead interactions influence the final predictions, thereby providing technical guidance for future improvement in clinical applications.

## 5   Conclusion and Perspectives

FetalHealthNet (FHNet) is presented as an encoder–decoder framework for high-fidelity fECG reconstruction from multi-lead aECG. By combining graph-based encoders (dynamic spatio-temporal connectivity with Chebyshev-GCN) and multi-scale attention (self-/cross-attention with MSTFE), the model captures inter-lead dependencies and preserves morphology under low SNR, achieving $R^2$ = 0.995 and RMSE = 0.0155 on representative benchmarks. Interpretability is supported by Integrated Gradients and spatial-attention maps that highlight time-resolved, lead-level contributions aligned with physiological expectations, facilitating clinical review.

Furthermore, through sophisticated Integrated Gradients analysis and comprehensive attention map visualization, FHNet provides unprecedented transparency in its decision-making architecture, systematically prioritizing physiologically critical leads while intelligently suppressing artifact-contaminated channels. This interpretability paradigm ensures seamless integration with clinical expertise and establishes robust foundations for evidence-based diagnostic protocols. By establishing an unprecedented benchmark that seamlessly integrates predictive excellence with comprehensive interpretability, FHNet fundamentally redefines clinician-AI collaboration while providing a groundbreaking methodological template with broad applicability across diverse physiological signal analysis domains in biomedical engineering and computational medicine.

## ACKNOWLEDGMENTS

The authors gratefully acknowledge the Office of Nevada Economic Development's Sports Innovation Initiative Grant at the University of Nevada, Las Vegas, for partial support of this work. We also extend our sincere gratitude to Mr. Daze Lu for his dedication and contribution to this work.